\documentclass[sigconf]{acmart} 

\usepackage{booktabs} 
\usepackage{tabularx}
\usepackage{multirow} 
\usepackage{balance}
\usepackage{makecell}
\usepackage{subfig} %
\usepackage{paralist}
\usepackage{fancyvrb}
\usepackage{xspace}

\newcommand{\q}[1]{\lq\lq{}{}#1\rq\rq{}{}}
\newcommand{\corona}{COVID-19}

\newcommand{\ie}{\textit{i.e.}\xspace}
\newcommand{\eg}{\textit{e.g.}\xspace}
\newcommand{\cf}{\textit{cf.}\xspace}
\newcommand{\etal}{\textit{et al.}\xspace}

\newcommand{\para}[1]{\smallskip\noindent\textbf{#1.}}

\copyrightyear{2020}
\acmYear{2020}
\setcopyright{acmcopyright}
\acmConference[CIKM '20]{Proceedings of the 29th ACM International Conference on Information and Knowledge Management}{October 19--23, 2020}{Virtual Event, Ireland}
\acmBooktitle{Proceedings of the 29th ACM International Conference on Information and Knowledge Management (CIKM '20), October 19--23, 2020, Virtual Event, Ireland}
\acmPrice{15.00}
\acmDOI{10.1145/3340531.3412765}
\acmISBN{978-1-4503-6859-9/20/10}

\begin{document}
\fancyhead{}

\title[]{TweetsCOV19 - A Knowledge Base of Semantically Annotated Tweets about the COVID-19 Pandemic}

\author{Dimitar Dimitrov$^{1}$, Erdal Baran$^{1}$, Pavlos Fafalios$^{2}$, Ran Yu$^{1}$, Xiaofei Zhu$^{3}$, Matth\"aus Zloch$^{1}$, and Stefan Dietze$^{1,4,5}$}
\affiliation{
 \institution{
 $^{1}$GESIS - Leibniz Institute for the Social Sciences, Cologne, Germany\\ 
 $^{2}$Institute of Computer Science, FORTH-ICS, Heraklion, Greece\\
 $^{3}$Chongqing University of Technology, Chongqing, China\\
 $^{4}$Heinrich-Heine-University D\"usseldorf, Germany\\
 $^{5}$ L3S Research Center, Hannover, Germany
 }
  \city{}
  \state{}
}
\email{
 {dimitar.dimitrov,erdal.baran,ran.yu, matthaeus.zloch,stefan.dietze}@gesis.org}
\email{fafalios@ics.forth.gr}
\email{zxf@cqut.edu.cn}

\begin{abstract}
Publicly available social media archives facilitate research in the social sciences and provide corpora for training and testing a wide range of machine learning and natural language processing methods. With respect to the recent outbreak of the Coronavirus disease 2019 (COVID-19), online discourse on Twitter reflects public opinion and perception related to the pandemic itself as well as mitigating measures and their societal impact. Understanding such discourse, its evolution, and interdependencies with real-world events or (mis)information can foster valuable insights. On the other hand, such corpora are crucial facilitators for computational methods addressing tasks such as sentiment analysis, event detection, or entity recognition. However, obtaining, archiving, and semantically annotating large amounts of tweets is costly. In this paper, we describe \textit{TweetsCOV19}, a publicly available knowledge base of currently more than 8 million tweets, spanning October 2019 - April 2020. Metadata about the tweets as well as extracted entities, hashtags, user mentions, sentiments, and URLs are exposed using established RDF/S vocabularies, providing an unprecedented knowledge base for a range of knowledge discovery tasks. Next to a description of the dataset and its extraction and annotation process, we present an initial analysis and use cases of the corpus.
\end{abstract}




\maketitle

\section{Introduction}
Social web platforms have emerged as a primary forum for online discourse. Such user-generated content can be seen as a comprehensive documentation of societal discourse of immense historical value for future generations~\cite{bruns2016twitter} and as an important resource for contemporary research. On the one hand, research in the computational social sciences relies on social media data to gain novel insights, for instance, about the spreading pattern of false claims on Twitter~\cite{vosoughi2018spread} or prevalent biases observable in online discourse~\cite{chakraborty2017makes}. On the other hand, computational methods at the intersection of natural language processing (NLP) and machine learning (ML) such as sentiment analysis~\cite{10.1145/3348445.3348466}, classification of news web pages, users or posts~\cite{popat2017truth}, or fake news detection~\cite{Tschiatschek:2018:FND:3184558.3188722} rely on social web corpora for training and evaluation.

Twitter specifically has been recognized as an important data source, facilitating research focused on insights or methods related to online discourse. In particular, during the recent COVID-19 pandemic, online discourse on Twitter has proved crucial to facilitate an understanding of the impact of the pandemic, implemented measures, societal attitudes and perceptions in this context and, most importantly, the interdependencies between public opinion and relevant political actions, policies, media events or scientific discoveries. Recent corpora include a multilingual dataset of COVID-19-TweetIDs~\cite{info:doi/10.2196/19273}  consisting of more than 129 million tweet IDs, or a tweet corpus with sentiment annotations released by Lamsal~\cite{781w-ef42-20}. Next to datasets focused on COVID-19 as a whole, datasets on other related topics have been created, for instance, about vaccines~\cite{muller2019crowdbreaks} covering sentiment-annotated tweets since June 2017 mentioning vaccine-related keywords.

However, given the legal and computational challenges involved in processing, reusing and publishing data crawled from Twitter, existing corpora usually consist of either raw metadata (such as tweet IDs, usernames, publishing dates)~\cite{huang_xiaolei_2020_3735015} or very limited and only partially precomputed features, such as georeferences~\cite{qazi_geocov19_sigspatial}. In addition, corpora tend to be tailored towards a technical audience, limiting reuse by non-technical research disciplines lacking the skills and infrastructure for large-scale data processing.

To facilitate a variety of multi-aspect data exploration scenarios, in prior work, we introduced \textit{TweetsKB}\footnote{\url{https://data.gesis.org/tweetskb}}---a pipeline for semantic annotation of tweets and a knowledge base of RDF data for more than 1.5 billion tweets spanning almost five years exposed using established vocabularies~\cite{fafalios2018tweetskb}.
Whereas entity-centric access and exploration methods are crucial to facilitate exploration of large Twitter archives, TweetsKB consistently applies W3C data sharing standards to publish a long-term Twitter archive in the form of an extensible and easy to access knowledge graph including metadata about tweets, user mentions, disambiguated entities, and sentiments. Building on the prior release of TweetsKB in 2018, this work provides the following contributions:
\begin{itemize}
    \item \textbf{Extension of TweetsKB.} Building on a continuous Twitter crawl and a parellelised annotation pipeline, we expand TweetsKB with data from April 2018 up to now, including additional metadata of about 486 million tweets, adding up to an unprecedented corpus of more than 63 billion triples describing more than 2 billion tweets starting from February 2013. To the best of our knowledge, TweetsKB is the largest publicly available Twitter archive and the only dataset consistently providing a knowledge graph of tweet metadata and precomputed features about entities and sentiments. Next to adding additional data based on our enrichment and data lifting pipeline (\cf Section \ref{sec:corpus}), we also extend both the applied schema and enrichment pipeline in order to include additional features (shared URLs).
    \item \textbf{Extraction and publishing of
    \textit{TweetsCOV19}, a knowledge graph of COVID-19-related online discourse}. Taking advantage of TweetsKB and related infrastructure, we extract TweetsCOV19, a unique corpus of COVID-19-related online discourse. By applying a well-designed seed list (\cf Section \ref{subsec:CovidGeneration}), we extract a TweetsKB subset spanning the period October 2019 - April 2020 and apply the same feature extraction and data publishing methods as for TweetsKB. This results in a dataset containing more than 270 million triples describing metadata for about 8.1 million tweets from 3.6 million Twitter users. Data is accessible as downloadable dumps following the N3 format and can be queried online through a dedicated, HTTP-accessible SPARQL endpoint. An easy to process \textit{tsv} file is provided in addition.
    \item \textbf{Initial descriptive data analysis, use cases, tasks and reuse.} Next to providing basic statistics about TweetsKB in general, we provide initial analysis and exploration of the TweetsCOV19 data (\cf Section \ref{subsec:covanalysis}) in order to facilitate an understanding and reuse of the dataset. In order to facilitate and document reuse and impact of the data, we introduce a number of use cases, discuss prior use (\cf Section \ref{sec:usecases}) of the data, for instance, to facilitate research in the social sciences, as well as additional machine learning and NLP tasks facilitated by TweetsCOV19. Among others, these include the task of predicting tweet virality, posed as a computation challenge (\cf Section~\ref{sec:analyticup}).
\end{itemize}

Given the fact that all Twitter corpora are prohibited from republishing actual tweet texts, precomputed features that reflect content and semantics of individual tweets, such as mentioned entities, hashtags, or URLs, together with expressed sentiments provide a unique foundation for studying online discourse and its evolution over time. To the best of our knowledge, TweetsCOV19 is the only COVID-19-related dataset available as public knowledge graph of tweets metadata and semantic annotations following established vocabularies and Web data sharing standards.

\section{Constructing a Knowledge Base of Twitter Discourse}
\label{sec:corpus}

\begin{table*}[]
    \caption{Descriptive statistics of TweetsKB and TweetsCOV19.}
    \vspace{-3mm}
    \centering
    \renewcommand{\arraystretch}{0.7}
    \setlength{\tabcolsep}{4.5pt}
    \begin{tabular}{l|r|r|c|r|r|c}
    \toprule
    & \multicolumn{3}{c}{TweetsKB} & \multicolumn{3}{c}{TweetsCOV19}  \\
      feature & total & unique & \begin{tabular}{@{}c@{}}ratio of tweets with  \\ at least one feature\end{tabular}  & total & unique & \begin{tabular}{@{}c@{}}ratio of tweets with  \\ at least one feature\end{tabular}  \\
     \midrule
       hashtags & 739,642,147 & 52,244,423 & 0.19 & 3,653,928 & 566,308 & 0.30 \\
       mentions & 1,072,723,250 & 116,499,222 & 0.35 & 5,363,449 & 1,251,963 & 0.40 \\
       entities & 2,575,861,358 & 1,919,083 & 0.58 & 11,537,537 &  331,307 & 0.70 \\
       non-neutral sentiment     & 1,047,840,159 & - & 0.54 & 4,478,603 & - & 0.55 \\
    \bottomrule
    \end{tabular}

    \label{tab:combinedstats}
    \vspace{-0.5em}
\end{table*}

Whereas the processing of TweetsCOV19, described in Section \ref{sec:tweetscov19}, builds on TweetsKB, here, we describe the construction process of TweetsKB as a general, large-scale knowledge base of Twitter discourse. Note that, next to updating the corpus with crawled data after the previous release, improvements were made to the processing pipeline for this release compared to the extraction process described in ~\cite{fafalios2018tweetskb}.

TweetsKB is a public RDF corpus containing a unique collection of more than 2 billion semantically-annotated tweets spanning more than 7 years (February 2013 - April 2020). Metadata about the tweets as well as extracted entities, sentiments, hashtags and user mentions are exposed using established RDF/S vocabularies, forming a large knowledge graph of tweet-related data and allowing the expression of structured (SPARQL) queries that satisfy complex/analytical information needs (\cf Section \ref{sec:usecases}). TweetsKB is generated through  the following steps:
(i) harvesting, 
(ii) filtering, 
(iii) cleaning,
(iv) semantic annotation and metadata extraction,
(vi) data lifting (using a dedicated RDF/S model).
Below we describe these steps, the availability of the extension of the TweetsKB dataset, and provide some descriptive statistics.

\para{Harvesting, filtering, cleaning} Tweets are continuously harvested through the public
Twitter streaming API since January 2013, accumulating more than 9.5 billion tweets up to now (May 2020).
While all data is being archived locally and on restricted servers, TweetsKB is based on the cleaned-up English-language subset. As part of the filtering step, we eliminate retweets and non-English tweets, reducing the number of tweets to about 2.3 billion tweets. In addition, we remove spam through a Multinomial Naive Bayes (MNB) classifier, trained on the HSpam dataset which has 94\% precision on spam labels~\cite{sedhai2015hspam14} removing an additional 10\% of tweets.

\para{Semantic annotation and metadata extraction} %
Adhering to the Twitter license terms, the text of each tweet is not republished itself but only tweet IDs which may be rehydrated for specific purposes. In addition, full-text is exploited for extracting and disambiguating mentioned entities (\textit{entity linking}), as well as for extracting the magnitude of the expressed positive and negative sentiments (\textit{sentiment analysis}). 
Relying on the experimental motivation and prior work in~\cite{fafalios2018tweetskb}, for \textit{entity linking}, we exploit Yahoo's Fast Entity Linker (FEL)~\cite{blanco2015fast}. 
FEL has shown particularly cost-efficient performance  on the task of linking entities from short texts to Wikipedia and is fast and lightweight, being well-suited to run over billions of tweets in a distributed manner.
We trained the FEL model using a Wikipedia dump of April 2020 and we set a confidence threshold of -3 which has been shown empirically to provide annotations of good quality (favoring precision). Using a current Wikipedia dump allows linking to COVID-19-related entities (\cf Section~\ref{sec:tweetscov19}). We also store the confidence score of each extracted entity to facilitate data consumers to set confidence scores which suit their use cases and requirements when working with our precomputed annotations. The quality of the entity annotations produced by FEL over tweets was evaluated in~\cite{fafalios2018tweetskb}, demonstrating high precision (86\%) and an overall satisfactory performance (F1 = 54\%). 

For sentiment analysis, we used SentiStrength~\cite{thelwall2012sentiment}, a robust and efficient tool for sentiment strength detection on social web data. SentiStrength assigns both a positive and a negative score to a short text, to account for both types of sentiments that can be expressed at the same time. The value of a positive (negative) sentiment ranges from +1 (-1) for no positive (no negative) to +5 (-5) for extremely positive (extremely negative).  
We provide an evaluation of the quality of sentiment annotations produced by SentiStrength over tweets in~\cite{fafalios2018tweetskb}, demonstrating a reasonable performance, in particular in distinguishing stronger sentiments. 

Entity and sentiment annotations are accompanied by the following metadata extracted from the tweets: 
\textit{tweet id}, 
\textit{post date}, 
\textit{username} (user who posted the tweet), 
\textit{favourite} and \textit{retweet count} (at the time of fetching the tweet),
\textit{hashtags} (words starting with \#), and 
\textit{user mentions} (words starting with @). 
Starting from April 2018, we also extract the \textit{URLs} included in the tweets. 
For ensuring data privacy, we anonymize usernames to ensure that tweets for particular users can be aggregated but users not identified.

\para{Data lifting} %
We generate RDF triples in the N3 format using the data model described in~\cite{fafalios2018tweetskb}, which exploits terms from established vocabularies, in particular
SIOC core ontology~\cite{breslin2006sioc},
ONYX~\cite{sanchez2016onyx}, 
schema.org~\cite{guha2016schema}, and
Open NEE Model~\cite{fafalios2015exploiting}.
The selection of vocabularies was based on the following objectives: (i) avoiding schema violations, (ii) enabling data interoperability through term reuse, (iii) having dereferenceable URIs, (iv) extensibility. During lifting, we normalize sentiment scores in the range $[0, 1]$ using the formula:  $score = (|sentimentValue| - 1) / 4)$.
For this release, we extended the data model described in~\cite{fafalios2018tweetskb} with one additional property (\textit{schema:citation}) which refers to a URL mentioned in the tweet. Given that roughly 21\% of tweets contain URLs, providing means to analyze shared URLs and Pay-Level-Domains (PLDs)\footnote{
The PLD is a sub-domain of a public top-level domain (like {\tt .com}), for which users usually pay for. For example, the PLD for {\tt www.example.com} would be {\tt example.com}.}  provides additional opportunities for a range of research questions and tasks, for instance, with respect to the spreading of misinformation.  

\para{Descriptive statistics and  data availability} TweetsKB currently contains approximately 62.23 billion triples describing online discourse on Twitter. Table~\ref{tab:combinedstats} summarizes descriptive statistics of the dataset (February 2013 - April 2020). About 19\% of the tweets contain at least one hashtag and 35\% at least one user mention. FEL extracted at least one entity for 58\% of the tweets, while  the  average  number  of  entities  per  tweet  is 1.26. About 46\% of the tweets have no sentiment, \ie, the score is zero for both the positive and the negative sentiment. Finally, we report a statistic calculated only for the TweetsKB extension starting from April 2018 for which we also extracted shared URLs, \ie, 21\% of the tweets from April 2018 to April 2020 contain at least one URL.

The full TweetsKB is available as N3 files (split by month) through the Zenodo data repository (DOI: 10.5281/zenodo.573852),\footnote{\url{https://zenodo.org/record/573852}} under a {\em Creative Commons Attribution 4.0} license. 
For demonstration purposes, we have also set up a public SPARQL endpoint, currently containing a subset of about 5\% of the dataset\footnote{\url{https://data.gesis.org/tweetskb/sparql} (Graph IRI: \url{http://data.gesis.org/tweetskb})}. Example queries and more information are available through {\em TweetsKB}'s home page.\footnote{\url{https://data.gesis.org/tweetskb}}
The source code used for triplifying the data is available as open source on GitHub\footnote{\label{fn:git}\url{https://github.com/iosifidisvasileios/AnnotatedTweets2RDF}}.

\section{The TweetsCOV19 dataset}
\label{sec:tweetscov19}

\begin{table*}[]
    \centering
        \caption{Top five matching keywords, user mentions, hashtags, and pay-level-domains of TweetsCOV19.}
        \vspace{-3mm}
        \renewcommand{\arraystretch}{0.7}
    \setlength{\tabcolsep}{4.5pt}
\begin{tabular}{lr|lr|lr|lr}
\toprule
keywords &  frequency    &  user mentions &  frequency & hashtags &  frequency & PLDs &  frequency  \\
\midrule
 ppe &    3,368,192 & realdonaldtrump &     41,839 & covid19 &     160,585 & twitter.com &     251,839 \\
coronavirus &    2,363,080 &      narendramodi &      13,039  & coronavirus &     148.317 &  youtube.com &      99,505\\
 covid &    2,308,054 &    pmoindia &      12,701 & covid\_19 &      27,049 & instagram.com &      50,846  \\
corona &    1,513,195 &    jaketapper &  9,836 & stayhome &      26,542 & nytimes.com &      30,892 \\
covid19 &    1,498,386 &      who &       9,776  & china &      23,602 &  theguardian.com &      26,737 \\
\bottomrule
\end{tabular}

    \label{tab:topkeywordsmentionshashtags}
    \vspace{-0.5em}
\end{table*}

In this section, we describe the extraction of the TweetsCOV19 dataset\footnote{\label{fn:hp}\url{https://data.gesis.org/tweetscov19}} --- a subset of TweetKB containing tweets related to COVID-19, which captures online discourse about various aspects of the pandemic and its societal impact. Applications and use of the data are described in greater detail in Section~\ref{sec:usecases}. We discuss the extensibility, sustainability, and maintenance of the dataset in Section~\ref{sec:sustainability} before providing a thorough comparison of TweetsCOV19 and related datasets in Section~\ref{sec:relatedwork}.

\subsection{Extraction Procedure \& Availability}
\label{subsec:CovidGeneration}
To extract the dataset, we compiled a seed list of 268 \corona-related keywords\footnote{\url{https://data.gesis.org/tweetscov19/keywords.txt}}. 
The seed list is an extension of the seed list\footnote{\url{https://github.com/echen102/COVID-19-TweetIDs/blob/master/keywords.txt}} of Chen \etal~\cite{info:doi/10.2196/19273} and allows a broader view on the societal discourse on \corona~in Twitter. We conducted full text filtering on the cleaned full-text of English tweets (\cf Section \ref{sec:corpus}) and retain all tweets containing at least one of the keywords in the seed list. We consider only original tweets and no retweets. Overall, our corpus contains 16,266,285 occurrences of matching seed keywords. We applied the same process to extract relevant metadata and semantically enrich each tweet as described in Section~\ref{sec:corpus}. To simplify analysis of the posted URLs, we resolved all shortened URLs.

The current state of the full dataset is available in two formats: (i) as a text file with tabular separated values (tsv) and (ii) as RDF triples in N3 format (\cf Section~\ref{sec:corpus}). 
The N3 version of the dataset consists of $274,451,101$ RDF triples accessible through a dedicated SPARQL-endpoint\footnote{\url{ https://data.gesis.org/tweetscov19/sparql} (Graph IRI: \url{ http://data.gesis.org/tweetscov19})} and as downloadable dumps\footnote{\url{https://zenodo.org/record/3871753}}. The source code used for triplifying the data is available as open source on GitHub\footnote{\label{fn:git}\url{https://github.com/iosifidisvasileios/AnnotatedTweets2RDF}}. All data is available under a {\em Creative Commons Attribution 4.0} license. New data will be incrementally added to the corpus in the future.

\subsection{Initial Data Analysis}
\label{subsec:covanalysis}
In this section, we present a preliminary and non-exhaustive analysis of the TweetsCOV19 dataset in order to facilitate an understanding of the data and captured features. The TweetsCOV19 dataset consists of 8,151,524 original tweets posted by 3,664,518 users captured during October 2019 - April 2020. Table~\ref{tab:combinedstats} shows descriptive statistics of TweetsCOV19. Given the applied FEL confidence threshold (-3 \cf Section~\ref{sec:corpus}), we note a high percentage of tweets containing entities (70\%). The relatively low number of unique entities, on the other hand, is most probably due to the topic-specific nature of the corpus. Comparing the ratio of tweets with at least one feature across TweetsKB and TweetsCOV19, we observe constantly higher numbers (at least 5\%) for all features, with non-natural sentiment being the sole exception (about 1\%).  Table~\ref{tab:topkeywordsmentionshashtags} shows the top five matching keywords, user mentions, hashtags, and PLDs in TweetsCOV19.
The most frequently matching keyword---\textit{ppe}---is the acronym for personal protective equipment such as face masks, eye protection, and gloves. Politicians, journalists, and health organizations are the most frequent user mentions, with \textit{@realdonaldtrump} being  by far the most frequently mentioned Twitter user, while the most used hashtags are \textit{\#covid19} and \textit{\#coronavirus}. Apart from URLs to Twitter and other social media platforms, URLs from PLDs of major news outlets appear to be most frequently shared. 

The TweetsCOV19 dataset contains 2,148,490 URLs from 1,645,394 distinct pay-level-domains. We observe that about 25\% of the tweets in TweetsCOV19 contain at least one URL, compared to 21\% in TweetsKB (\cf Section \ref{sec:corpus}). The higher proportion of URLs seems intuitive given that for emerging topics such as COVID-19, sharing informational resources is one of the primary motivations.

\begin{figure*}
    \centering
    \subfloat[][\#coronavirus vs. \#covid19]{\includegraphics[width=0.5\textwidth]{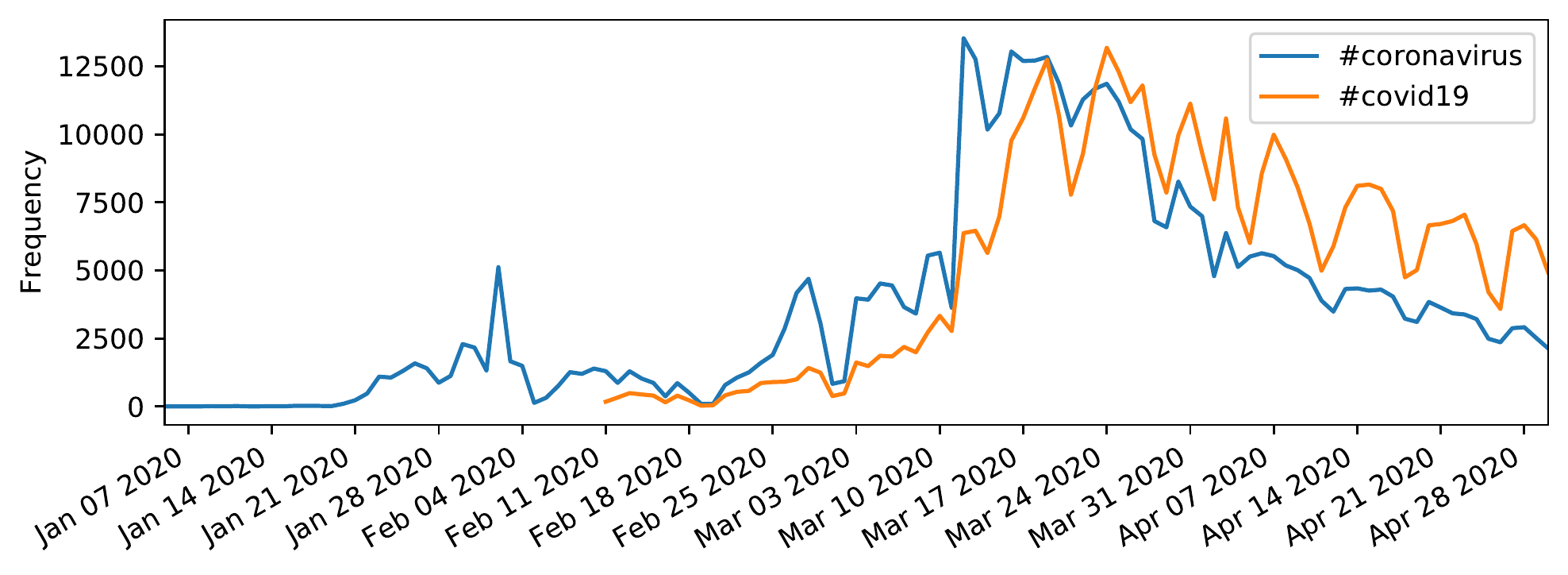}\label{fig:covidvscorona}}
    \subfloat[][\#hydroxychloroquine vs. \#vaccine]{\includegraphics[width=0.49\textwidth]{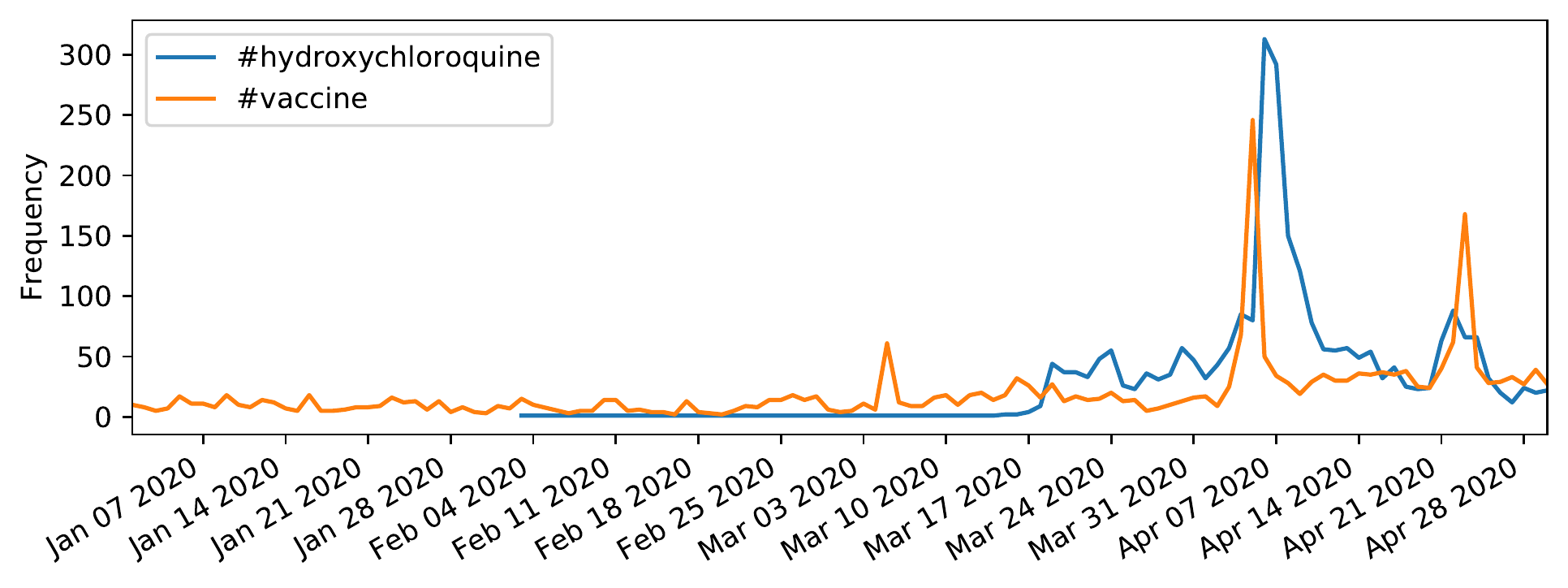}\label{fig:vaccinevshydro}}
    
    \caption{Hashtag usage over time. The figure shows a comparison of hashtag popularity over time for \protect\subref{fig:covidvscorona} the two most popular hashtags \textit{\#coronavirus} and \textit{\#covid19}, and for \protect\subref{fig:vaccinevshydro} \textit{\#hydroxychloroquine} vs. \textit{\#vaccine}.
    }
    \label{fig:coronavscovid19-popularity}
    \vspace{-1em}
\end{figure*}

Although the TweetsCOV19 dataset contains data from October 2019 to April 2020, our next analysis concentrates on the period from January to April 2020, where the topic starts dominating social media. Figure~\ref{fig:coronavscovid19-popularity} presents a comparison of hashtag popularity over time for \textit{\#coronavirus} vs. \textit{\#covid19} and \textit{\#hydroxychloroquine} vs. \textit{\#vaccine}. The hashtag \textit{\#coronavirus} is present for the whole period and shows a small initial peak just before the emergence of \textit{\#covid19} in the beginning of February 2020. %
While \textit{\#vaccine} is a topic that has been receiving attention on social media even before the COVID-19 crisis, \textit{\#hydroxychloroquine} gained first popularity as a possible drug for treating COVID-19 patients. Nevertheless, mentions of both terms seem to be strongly correlated. %
As desired, COVID-19 is present not only with respect to hashtags but also with respect to the semantic annotations of the data.
Table~\ref{tab:entitiesovertimethreshold-2} shows the top five most frequently recognized entities per month from January to April 2020. The entity \textit{Coronavirus\_disease\_2019} experiences a drastic boost in March and April 2020 and is overall the most frequent in TweetsCOV19.

\begin{figure*}[t]
\centering
\subfloat[][Donald Trump]{\includegraphics[width=0.5\textwidth]{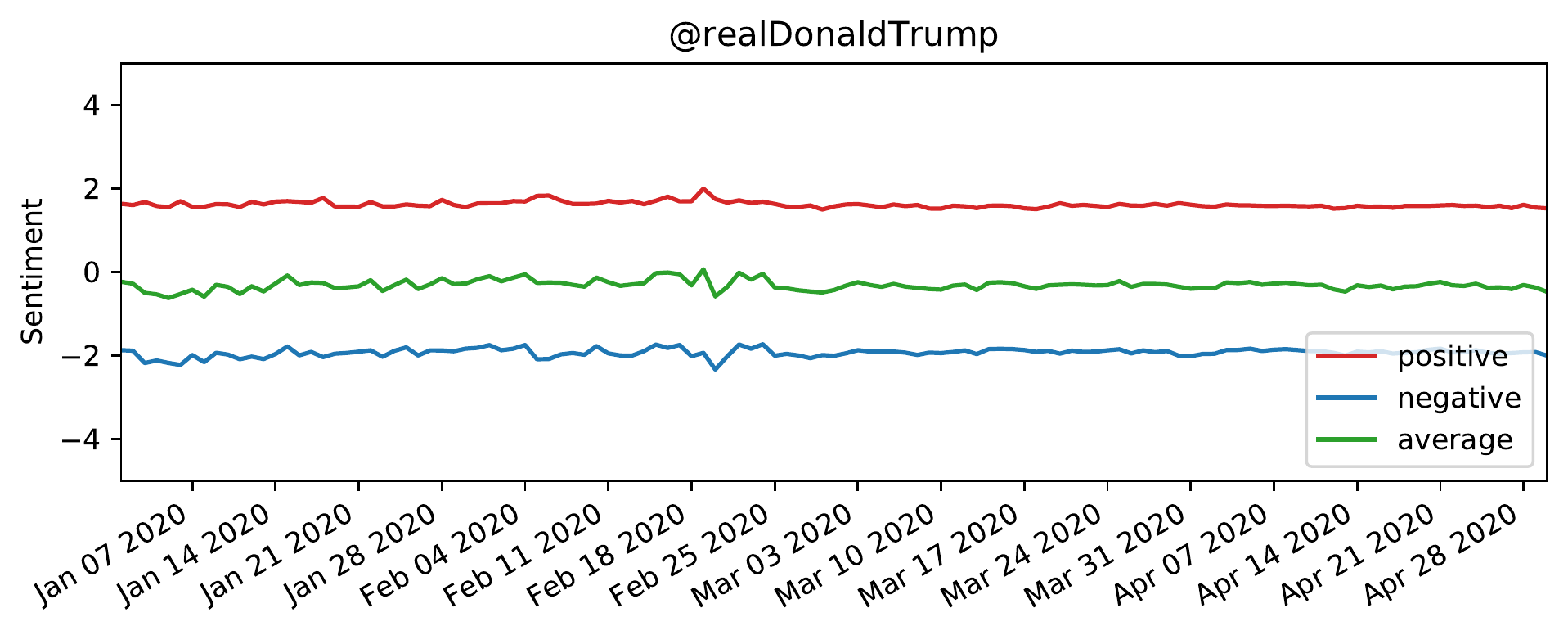}\label{fig:realDonaldTrump-mention-sentimen}}
\subfloat[][WHO]{\includegraphics[width=0.5\textwidth]{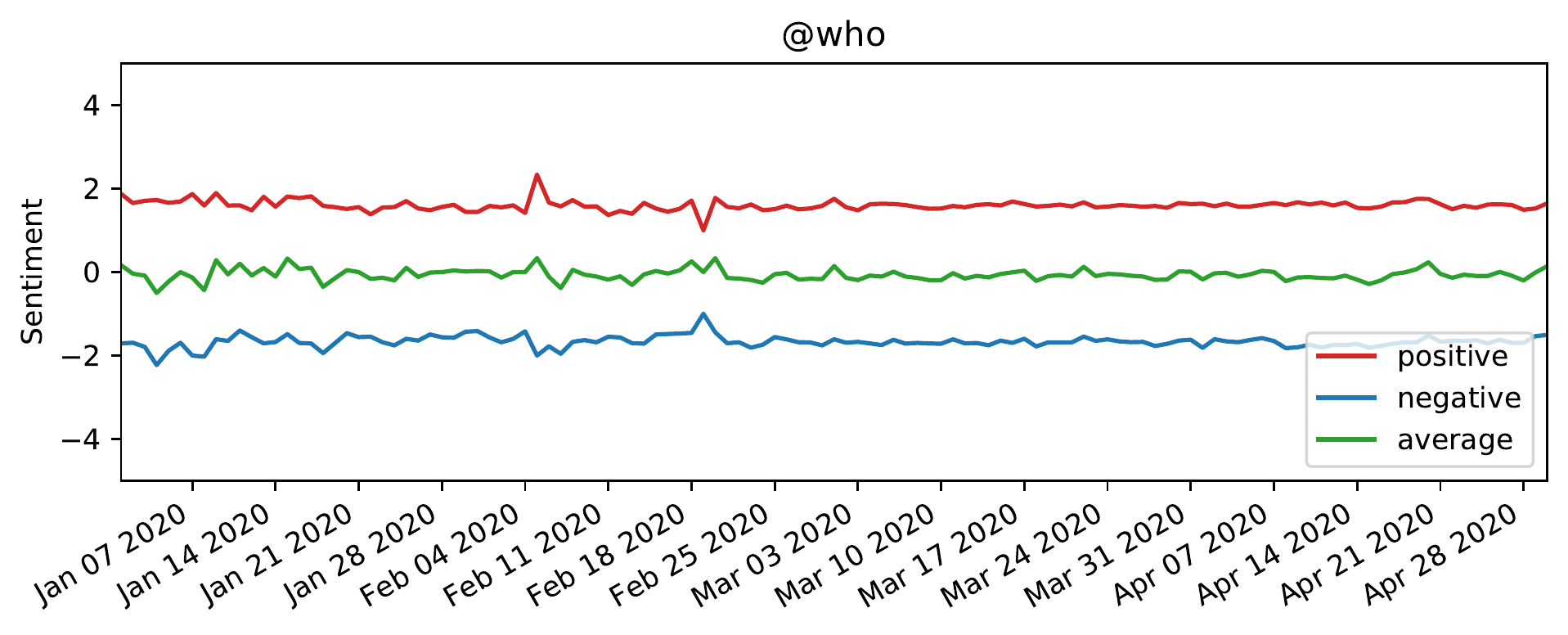}\label{fig:who-mention-sentiment}}

\subfloat[][Breitbart]{\includegraphics[width=0.5\textwidth]{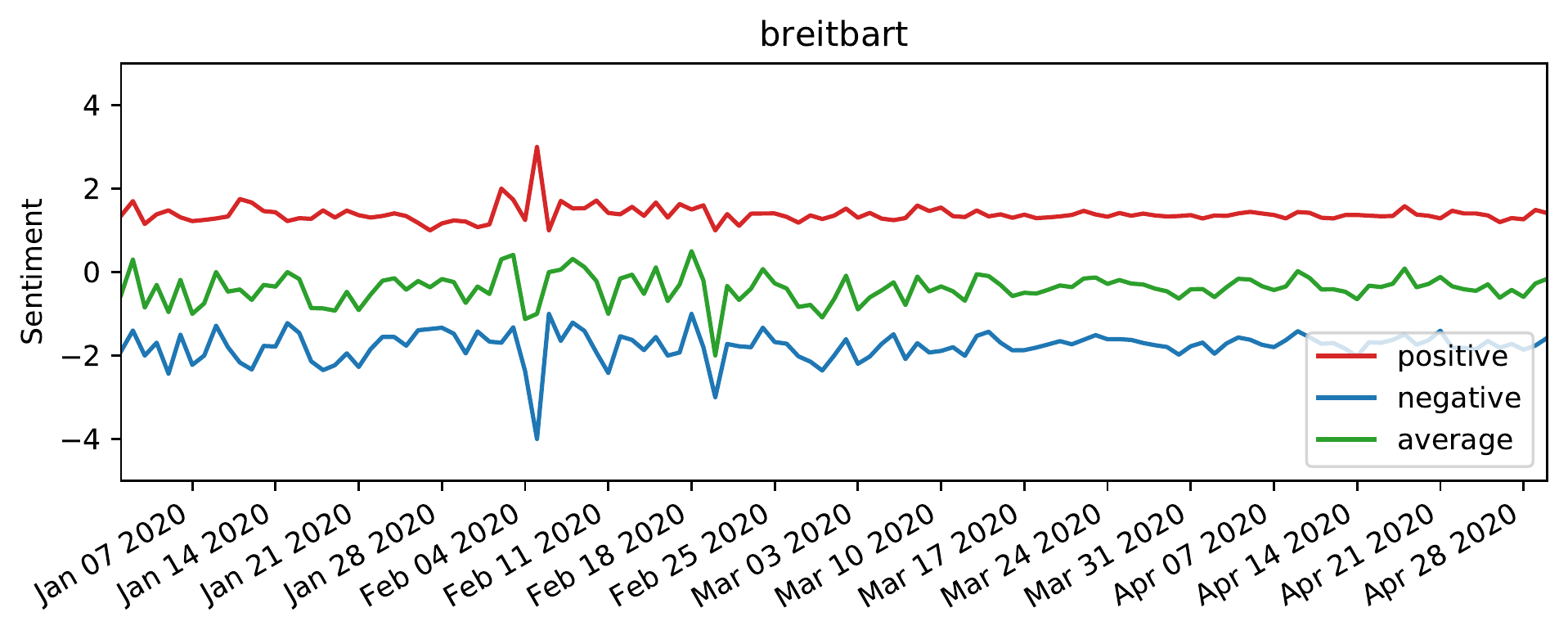}\label{fig:breitbart-pld-sentiment}}
\subfloat[][CNN]{\includegraphics[width=0.5\textwidth]{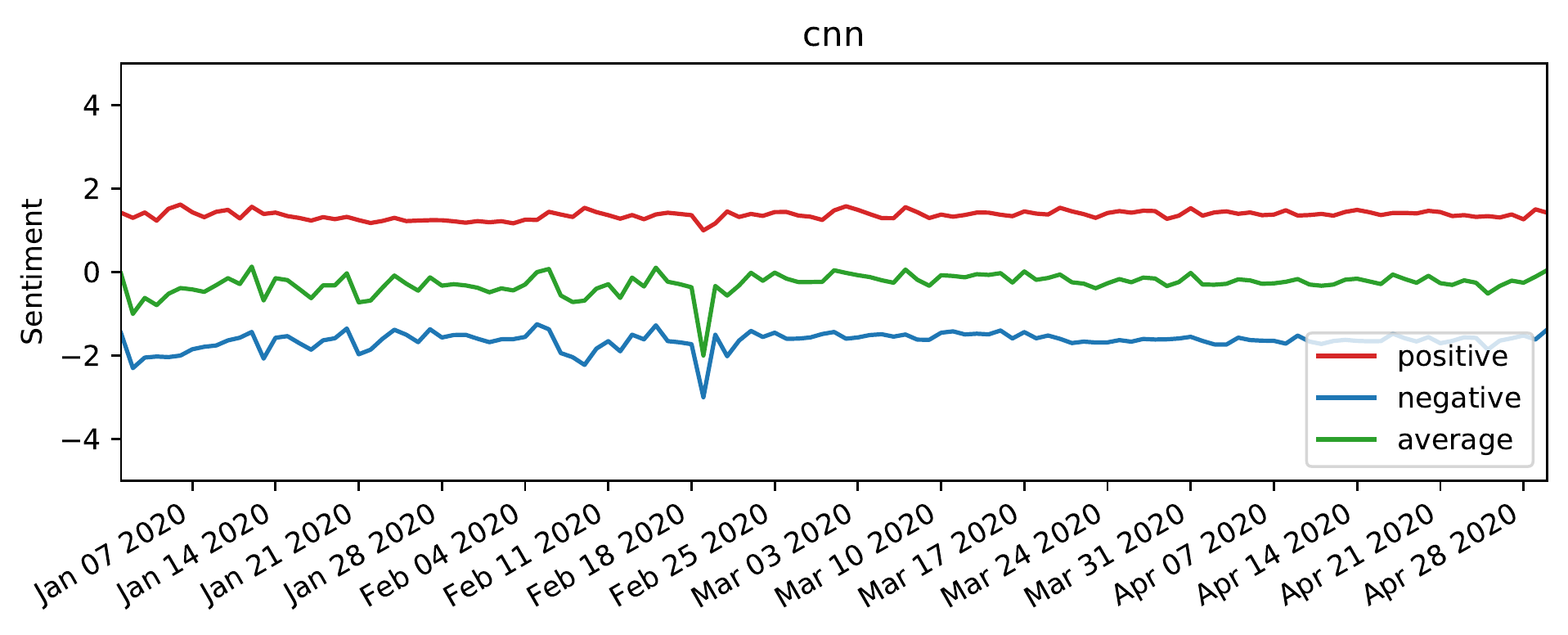}\label{fig:cnn-pld-sentiment}}

\caption{\textbf{Sentiment over time. The figure shows the sentiment of tweets mentioning \protect\subref{fig:realDonaldTrump-mention-sentimen} Donald Trump and \protect\subref{fig:who-mention-sentiment} WHO, and containing URLs to \protect\subref{fig:breitbart-pld-sentiment} Breitbart---a politically far-right-wing associated news media---and \protect\subref{fig:cnn-pld-sentiment} CNN---a left-wing associated media.} %
}
\label{fig:urlmentionssentiment}
\vspace{-1em}
\end{figure*}

In the context of the pandemic, Figure~\ref{fig:urlmentionssentiment} illustrates the sentiments of tweets containing relevant user mentions and URLs to news media outlets. %
For all tweets on a given day, the figure shows their average positive (red), average negative (blue) sentiment, and the sum of positive and negative sentiments for a tweet averaged overall tweet for the day (green).   
While a systematic detection and interpretation of events is out of the scope of this paper, the fluctuation of the sentiment towards Donald Trump (\cf Figure~\ref{fig:urlmentionssentiment}\subref{fig:realDonaldTrump-mention-sentimen})  may be better understood in the context of an excerpt from the timeline of major events as compiled by the Washington Post\footnote{\url{https://www.washingtonpost.com/politics/2020/04/20/what-trump-did-about-coronavirus-february}}. The timeline highlights Trump's mention of the COVID-19 outbreak in his state of the union address (February 4, 2020) and the United States Department of Health and Human Services (HHS) announcing its first efforts to rapidly develop a COVID-19 vaccine in cooperation with pharmaceutical industry representatives (February 18, 2020).
    
Another view on these events is provided by the sentiment of tweets containing URLs to Breitbart---a politically far-right-wing associated news media (\cf  Figure~\ref{fig:urlmentionssentiment}\subref{fig:breitbart-pld-sentiment}) and CNN---a left-wing associated media (\cf Figure~\ref{fig:urlmentionssentiment}\subref{fig:cnn-pld-sentiment}). We observe the strongest fluctuations in the sentiment, with the biggest divergence of sentiment to tweets sharing links to Breitbart articles in the week of Trump's State of the Union address (February 4, 2020). The strongly diverging spikes, both positivity and negativity increase at the same time, suggest a possible controversiality and polarization in Breitbart's coverage of topics related to the address. We observe the opposite converging pattern around February 22, 2020, for the sentiment of tweets mentioning the World Health Organization (WHO) (\cf Figure~\ref{fig:urlmentionssentiment}\subref{fig:who-mention-sentiment}).
The aspects associated with those sentiments patterns represent a promising direction for future investigations.

\begin{table*}[]
\vspace{-2em}
    \centering
   \caption{Entities over time. The table shows the top five entities (confidence score -2) and their frequency per month in the TweetsCOV19 dataset since the beginning of 2020. COVID-19* is used as shortcut for the entity \textit{Coronavirus\_disease\_2019}.}
   \vspace{-3mm}
   \renewcommand{\arraystretch}{0.7}
    \setlength{\tabcolsep}{4pt}
\begin{tabular}{lr|lr|lr|lr}
\toprule
\multicolumn{2}{c}{January 2020}      & \multicolumn{2}{c}{February 2020}     & \multicolumn{2}{c}{March 2020}             & \multicolumn{2}{c}{April 2020}     \\
entity             & frequency  & entity                    & frequency & entity                     & frequency & entity                   & frequency \\
\midrule
Wuhan              & 10,147     & Wuhan                     & 10,494  & COVID-19*   & 178,396 & COVID-19* & 200,342 \\
Iran               &  5,905     & COVID-19*  &  4,999  & Social\_distancing            &  66,176 & Social\_distancing          &  52,323 \\
BTS                &  5,014     & BTS                       &  4,513  & Italy                        &  22,164 & India                      &  18,992 \\
What's\_Happening!! &  4,899     & What's\_Happening!!        &  3,431  & Wuhan                        &  16,804 & Hydroxychloroquine         &  15,820 \\
Twitter            &  4,105     & Twitter                   &  3,351  & India                        &  15,822 & Wuhan                      &  14,478 \\
\bottomrule
\end{tabular}
    \label{tab:entitiesovertimethreshold-2}
    \vspace{-1em}
\end{table*}

\section{Use Cases \& Impact}
\label{sec:usecases}
This section describes usage and use cases of TweetsCOV19. 

\subsection{Exploitation Scenarios \& Example Queries}
Next to downloading the dumps, one can directly explore the full dataset or develop applications that make use the data through HTTP requests and the SPARQL endpoint, \eg, for retrieving specific data of interest, or for offering a user friendly interface on top of the endpoint, \eg, one similar to~\cite{gasquet2019exploring}.\footnote{Offering such a user-friendly interface is beyond the scope of this paper but considered future work.} In this section, we introduce a few basic queries to illustrate the use of the endpoint.

\begin{figure}[th]
\centering
\scriptsize
\begin{Verbatim}[frame=lines,numbers=left,numbersep=1pt]
select ?otherEntity (count(?tweet) as ?count) where {
  ?tweet schema:mentions ?entity ; dc:created ?date 
                FILTER(year(?date) = 2020 and month(?date) = 4) .
  ?entity a nee:Entity ; nee:hasMatchedURI dbr:Donald_Trump .
  ?tweet schema:mentions ?entity2 . 
  ?entity2 a nee:Entity ; nee:hasMatchedURI ?otherEntity 
                FILTER (?otherEntity != dbr:Donald_Trump)
} group by ?otherEntity order by desc(?count) LIMIT 5
\end{Verbatim}
\vspace{-2mm}
\caption{Top entities co-occurring with the entity \textit{Donald\_Trump} during April 2020.}
\label{fig:sparqlEx1}
\vspace{-1em}
\end{figure}

\begin{figure}[th]
\centering
\scriptsize
\begin{Verbatim}[frame=lines,numbers=left,numbersep=1pt]
select (day(?date) as ?day) (count(?tweet) as ?count) where {
  ?tweet schema:mentions ?entity ; dc:created ?date 
              FILTER(year(?date) = 2020 and month(?date) = 4) .
  ?entity a nee:Entity ; nee:hasMatchedURI dbr:Hydroxychloroquine .
} group by day(?date) order by day(?date)
\end{Verbatim}
\vspace{-2mm}
\caption{Number of tweets per day in April 2020 mentioning the entity \textit{Hydroxychloroquine}.}
\label{fig:sparqlEx2}
\vspace{-1em}
\end{figure}

\begin{figure}[th]
\centering
\scriptsize
\begin{Verbatim}[frame=lines,numbers=left,numbersep=1pt]
select ?url (count(?tweet) as ?count) where {
  ?tweet schema:mentions ?entity ; dc:created ?date 
    FILTER(year(?date)=2020 and month(?date)=4 and (day(?date)=6 || day(?date)=7)) .
  ?entity a nee:Entity ; nee:hasMatchedURI dbr:Hydroxychloroquine .
  ?tweet schema:citation ?url 
} group by ?url order by desc(?count)
\end{Verbatim}
\vspace{-2mm}
\caption{Top URLs mentioned in tweets of April 6, 2020 together with the entity \textit{Hydroxychloroquine}.}
\label{fig:sparqlEx3}
\vspace{-1em}
\end{figure}

Consider, for instance, that a user wants to investigate online discussions around the president of the United States in a specific time period, \eg, April 2020. The SPARQL query in Figure \ref{fig:sparqlEx1} retrieves the top five entities co-occurring with the entity \textit{Donald Trump} in tweets of April 2020 (\lq{}top\rq{} in terms of number of tweets mentioning the entities). The query returns the following five entities: 
\textit{China} (1,103 tweets), 
\textit{Coronavirus\_disease\_2019} (1,042 tweets), 
\textit{Hydroxychloroquine} (703 tweets), 
\textit{Disinfectant} (436 tweets), 
\textit{President\_of\_the\_United\_States} (369 tweets).
Individual entities of interest may be explored further. For instance, the SPARQL query in Figure \ref{fig:sparqlEx2} retrieves the number of tweets per day in April 2020 mentioning the entity \textit{Hydroxychloroquine}.
The results show a very high increment on April 6 and April 7, 2020 (from around 500 tweets before April 6 to 2,324 in April 6 and 2,105 in April 7), suggesting that some significant event related to this entity took place during this period. To explore this further, the SPARQL query in Figure \ref{fig:sparqlEx3} retrieves the top URLs included in tweets from April 6 and April 7, 2020, that mention the entity \textit{Hydroxychloroquine}. Headlines of the top three URLs are: 
\textit{\q{Detroit rep says hydroxychloroquine, Trump helped save her life amid COVID-19 fight}}\footnote{\url{https://www.freep.com/story/news/local/michigan/detroit/2020/04/06/democrat-karen-whitsett-coronavirus-hydroxychloroquine-trump/2955430001}} (54 tweets), 
\textit{\q{Trump's Aggressive Advocacy of Malaria Drug for Treating Coronavirus Divides Medical Community}}\footnote{
\url{https://www.nytimes.com/2020/04/06/us/politics/coronavirus-trump-malaria-drug.html}} (53 tweets), and
\textit{\q{Scoop: Inside the epic White House fight over hydroxychloroquine}}\footnote{\url{https://www.axios.com/coronavirus-hydroxychloroquine-white-house-01306286-0bbc-4042-9bfe-890413c6220d.html}} (30 tweets).
These news articles provide indicators about the event during April 6 - April 7, 2020, triggering a spike in popularity for \textit{Hydroxychloroquine}. 

\subsection{Ground Truth Data for ML and NLP Models}
\label{sec:analyticup}
From a methodological perspective, the corpus can serve as training/testing data for building and evaluating ML and NLP models. TweetsCOV19 can foster NLP research in different ways. For example, for each tweet, the dataset offers posted URLs and can thus provide ground truth data for building citation discovery models, \ie, predicting the URLs (PLDs) shared in a tweet. In cases where the ground truth is not directly provided by TweetsCOV19, the data can still be useful, \eg, tweets containing user mentions and entities can be easily identified for further annotation as part of crowd sourcing-based ground truth creation. In this way, the dataset can support building NLP models for aspect-based sentiment analysis, stance detection for political claims or fake news identification. Current work, for instance, is concerned with computing stances of tweets towards claims, such as the ones public in \textit{ClaimsKG}\footnote{\url{https://data.gesis.org/claimskg}}\cite{Tchech2019claimskg} and explicitly capture stances as metadata.

Furthermore, the dataset is a resource for building and evaluating machine learning models concerned with human online behavior. Predicting tweet virality is a concrete example where TweetsCOV19 is currently in use and provides the data for the \textit{COVID-19 Retweet Prediction Challenge}\footnote{\url{http://data.gesis.org/covid19challenge}} part of the CIKM 2020 AnalytiCup\footnote{\label{fn:analyticup}\url{https://cikm2020.org/analyticup}}. The goal of this challenge is to predict COVID-19-related tweets' virality in terms of the number of their retweets. Retweeting---re-posting original content without any change---is a popular function in online social networks and amplifies the spread of original messages. Understanding retweet behavior is useful and has many practical applications, \eg (political) audience design and marketing~\cite{stieglitz2012political,kim2014brand}, tracking (fake) news and misinformation~\cite{vosoughi2018spread,lumezanu2012bias}, social event detection~\cite{gupta2012predicting}. In particular, when designing campaigns of high societal impact and relevance, when handling communication through emergencies such as hurricane warnings~\cite{kogan2015think} and health-related campaigns about breast cancer screening~\cite{chung2017retweeting}, being able to predict future popularity of tweets is crucial. TweetsCOV19 can shape the understanding of such processes through the tweets metadata and semantic annotations it offers.

\subsection{Interdisciplinary Usage \& Impact}
The TweetsKB and TweetsCOV19 datasets are currently used to support interdisciplinary research in various fields. TweetsKB is currently used to shape the understanding of solidarity discourse in the context of migration, \eg, as part of the SOLDISK project\footnote{\url{https://www.uni-hildesheim.de/soldisk/en/project-description}}. In addition, ongoing joint work with media and communication studies researchers\footnote{\url{https://www.phil-fak.uni-duesseldorf.de/en/kmw/professur-i-prof-dr-frank-marcinkowski/research-areas}} uses TweetsKB to investigate the societal impact of the ongoing COVID-19 pandemic and most importantly, acceptance and trust for mitigating measures, the individual risk assessment and the impact of specific media events or information campaigns on related discourse and solidarity within society. In this context, in particular the impact of misinformation on solidarity and attitudes is being explored, taking advantage of the provided metadata together with additional metadata such as shared URLs and claims conveyed as part of these. Additional use cases are the joint exploration of means to extract statistically representative data for federal statistical agencies such as DESTATIS\footnote{\url{https://www.destatis.de/EN}} as a way to complement traditional data gathering instruments, such as survey programmes, which are not well-suited to capture societal discourse or dynamic interdependencies.

Among the lessons learned so far is that, despite all data preprocessing and enrichment aimed at simplifying reuse and interpretation of the data, data consumers tend to depend on support from computer and data scientists to handle and analyze the data. While in some cases, the critical issue is handling data at such a scale, in other cases, interpreting serialization formats (such as JSON or N3) or vocabularies poses challenges for users. Additionally, data quality problems related to the underlying raw data and preprocessed features call for highly collaborative projects where expertise in data characteristics and computational methods contributes to addressing higher-level research questions.

\section{Sustainability, Maintenance \& Extensibility}
\label{sec:sustainability}
With respect to ensuring long-term sustainability, two aspects are of crucial importance: (i) maintenance and sustainability of the corpus and enrichment pipeline, and (ii) maintenance of a user base and network. In order to ensure long-term sustainability, GESIS as research data infrastructure organisation exploits its technical expertise in hosting robust research data services has taken over the TweetsKB corpus with this recent update and hosts and maintains both TweetsKB and TweetsCOV19. Maintenance of the corpus will be facilitated through the continuous process of crawling 1\% of all tweets (running since January 2013) through the public Twitter API. In order to cater for downtimes and ensure that historic data is available for all time periods, redundant crawlers have been set up since March 2019. Storage of raw API output is currently handled through both, secure local GESIS storage services as well as the HDFS cluster at L3S Research Center.
 
The annotation and triplification process (\cf Section \ref{sec:corpus}) will be periodically repeated in order to incrementally expand the TweetsKB corpus and ensure its currentness, one of the requirements for many of the envisaged use cases of the dataset. While this will permanently increase the population of the dataset, the schema itself is extensible and facilitates the enrichment of tweets with additional information. For instance, information about the users involved in particular interactions (retweets, likes) or additional information about involved entities or references/URLs can be included. Next to facilitating the reuse of TweetsKB itself, we also publish the source code used to semantically annotate and triplify the data (\cf Footnote \ref{fn:git}), to enable third parties to establish and share similar corpora, \eg, focused topic-specific Twitter crawls. By following established W3C principles for data sharing and through the use of persistent URIs, both the schema as well as the corpus itself can be extended and linked. TweetsCOV19, in particular, will be updated continuously with the next release scheduled together with the submission deadline of the \textit{COVID-19 Retweet Prediction Challenge}. This allows us to utilize data from the period since this release as testing data for challenge participants.

A user base emerged gradually throughout the past years, most importantly, through enabling non-computer scientists (\eg social scientists) to interact and analyze the data (\cf Section \ref{sec:usecases}). Additionally, the corpus will be further advertised through interdisciplinary networks like the Web Science Trust\footnote{\url{http://www.webscience.org}}. Whereas the use of Zenodo for depositing the dataset, as well as its registration at {\tt datahub.ckan.io}, makes it citable and findable, we are currently exploring additional means, \eg, GESIS-hosted research data portals and registries to further publish and disseminate the dataset or particular subsets.

\section{Related Work}
\label{sec:relatedwork}
Several Twitter-related datasets similar to TweetsKB have emerged to support different fields such as machine learning and natural language processing. Compared to TweetsKB, some datasets contain only information filtered from raw Twitter stream data, \eg, to extract subsets of relevance to particular events\footnote{\url{https://digital.library.unt.edu/ark:/67531/metadc1259406}}. Others include annotations, such as mentioned entities~\cite{fafalios2018tweetskb}, or manually curated labels, \eg, sentiments\footnote{\url{https://data.world/crowdflower/weather-sentiment}}.

Since the pandemic started around January 2020, several datasets capturing COVID-19 discussions on Twitter have been released for academic use, including a new COVID-19-dedicated stream API by Twitter. The COVID-19 streaming API\footnote{\url{https://developer.twitter.com/en/docs/labs/covid19-stream}} returns tweets filtered based on 590 COVID-19-related keywords and hashtags (snapshot of terms on May 13, 2020) in the legacy enriched native response format\footnote{\url{https://developer.twitter.com/en/docs/tweets/data-dictionary/overview/tweet-object}}. We provide a summary of all relevant COVID-19-related datasets we found by the time of this study (\ie, May 20, 2020) at the TweetsCOV19 home page (\cf Footnote~\ref{fn:hp}). 
For the most of these datasets, tweets are harvested and filtered from the Twitter stream based on mentions of COVID-19-related keywords and hashtags~\cite{banda_juan_m_2020_3831406,huang_xiaolei_2020_3735015,DVN/LW0BTB_2020,qazi_geocov19_sigspatial,muller2019crowdbreaks,muller2019crowdbreaks}. The number of keywords and hashtags range from 3~\cite{DVN/LW0BTB_2020} to 800~\cite{qazi_geocov19_sigspatial}. Some of datasets further apply language~\cite{781w-ef42-20,alqurashi2020large,gao2020naist,0wf0-0792-20} or geo-location filters~\cite{fpsb-jz61-20}. Other datasets focus on specific languages. For example, ArCOV-19~\cite{haouari2020texttt} contains tweets returned by the Twitter standard search API\footnote{\url{https://developer.twitter.com/en/docs/tweets/search/api-reference/get-search-tweets}} when using COVID-19-related keywords (\eg, corona) as queries and written in Arabic, whereas others include only tweets in Turkish~\cite{0wf0-0792-20}. Most of the datasets are being updated regularly. The number of tweets contained varies per dataset and range from 747,599~\cite{haouari2020texttt} to over 524 million~\cite{qazi_geocov19_sigspatial} by the time of this study (\ie, May 20, 2020). The filtering criteria (\eg, keywords and selected user accounts) of all datasets are transparent. All datasets are available as \textit{csv}, \textit{tsv}, \textit{json} or plain text files for downloading. Each dataset provides the tweet IDs that can be used to rehydrate tweets, \ie, to acquire actual tweet content of the tweets. Some datasets also contain further information of tweets such as the publishing time~\cite{banda_juan_m_2020_3831406}, usernames~\cite{huang_xiaolei_2020_3735015,qazi_geocov19_sigspatial,gao2020naist,0wf0-0792-20}, geo-location~\cite{fpsb-jz61-20,qazi_geocov19_sigspatial,muller2019crowdbreaks} and retweet information~\cite{0wf0-0792-20}. A few datasets offer automatic annotations such as frequent used terms~\cite{banda_juan_m_2020_3831406}, sentiment scores per tweet~\cite{781w-ef42-20}, geo-location inferred from tweets~\cite{huang_xiaolei_2020_3735015} or places mentioned~\cite{qazi_geocov19_sigspatial,muller2019crowdbreaks}. The starting date of data collection varies, with the earliest available dataset providing data starting from January 1, 2020~\cite{alqurashi2020large}. The Vaccine Sentiment Tracking dataset~\cite{muller2019crowdbreaks} which is intended for sentiment analysis on vaccine-related topics even dates back to June 29, 2017. 

TweetsCOV19 differs from existing datasets as: (i) it is extracted from a permanent crawl (TweetsKB) spanning more than seven years -- facilitating to trace keywords and topics over extended periods of time, (ii) it has rich semantic annotations -- entities, sentiment scores, and URLs mentioned in tweets, (iii) the data is published following FAIR/W3C standards and can be accessed in various ways -- downloadable data dump as tab separated text files, RDF triples in N3 format, and a live SPARQL-endpoint. In particular, given that legal constraints prevent the republication of actual tweet text, precomputed features that reflect the semantics of tweets are both a distinctive feature of our dataset and a crucial requirement for efficiently analyzing online discourse.

\section{Conclusions}
As part of this work, we have performed a significant update of the TweetsKB dataset and pipeline. The most important part of the pipeline update is the usage of the April 2020 Wikipedia dump to perform entity linking, \ie, to enable linking to entities related to the ongoing COVID-19 pandemic. We have also introduced TweetsCOV19, a knowledge graph of Twitter discourse on COVID-19 and its societal impact. %
The corpus facilitates the exploration and analysis of online discourse even without costly feature computation or rehydration of tweets. Additionally, we have introduced several use cases from various disciplines, currently exploiting the corpus to derive insights and evaluate computational methods for various tasks.

Future work will take advantage of the extensible knowledge graph nature of the corpus to incrementally add further contextual information. TweetsKB and TweetsCOV19 can be extended by computing stances towards claims taking advantage of ClaimsKG~\cite{Tchech2019claimskg}, and by classifying tweets or users based on their shared URLs and claims allowing to detect misinformation.

\begin{acks}
We want to thank our colleagues at L3S Research Center, Hanover, Germany, and Humboldt University, Berlin, Germany, who initialized and are currently running the long-term Twitter crawl underlying TweetsKB and TweetsCOV19. We also thank the reviewers for their valuable comments and suggestions.
\end{acks}

\bibliographystyle{ACM-Reference-Format}
\balance
\bibliography{bibliography}

\end{document}